\documentclass{sig-alternate-05-2015}
\usepackage{algorithm}
\usepackage{algorithmicx}
\usepackage{algpseudocode}
\usepackage{graphicx}
\usepackage{tabularx}
\usepackage{multirow}
\usepackage{epstopdf}
\usepackage{etoolbox}

\patchcmd{\maketitle}{\@copyrightspace}{}{}{}
\begin{document}





%
\title{Audio Event Detection using Weakly Labeled Data}

\numberofauthors{2} 
%
\author{
%
%
\alignauthor
Anurag Kumar\\
       \affaddr{Language Technologies Institute}\\
       \affaddr{Carnegie Mellon University}\\
       \affaddr{Pittsburgh, USA}\\
       \email{alnu@andrew.cmu.edu}
\alignauthor
Bhiksha Raj\\
       \affaddr{Language Technologies Institute}\\
       \affaddr{Carnegie Mellon University}\\
       \affaddr{Pittsburgh, USA}\\
       \email{bhiksha@cs.cmu.edu}
}

\maketitle
\begin{abstract}
Acoustic event detection is essential for content analysis and description of multimedia recordings. The majority of current literature on the topic learns the detectors through fully-supervised techniques employing \emph{strongly} labeled data. However, the labels available for majority of multimedia data are generally weak and do not provide sufficient detail for such methods to be employed. In this paper we propose a framework for learning acoustic event detectors using only weakly labeled data.  We first show that audio event detection using weak labels can be formulated as an Multiple Instance Learning problem. We then suggest two frameworks for solving multiple-instance learning, one based on support vector machines, and the other on neural networks.  The proposed methods can help in removing the time consuming and expensive process of manually annotating data to facilitate fully supervised learning. Moreover, it can not only detect events in a recording but can also provide temporal locations of events in the recording. This helps in obtaining a complete description of the recording and is notable since temporal information was never known in the first place in weakly labeled data.
\end{abstract}
%

\vspace{-0.10in}
\section{Introduction}
The amount of consumer-generated multimedia data on the internet has grown almost exponentially in recent times. One popular multimedia upload site, YouTube, reported about a year ago  that 300 hours of multimedia recordings are uploaded on it every {\em minute} \cite{youtube}. There are several such sites on the internet today, each of which attracts similarly large amounts of data. The recordings are largely unannotated; descriptions if any are limited to simple high-level metadata such as the author, or a brief legend indicating the overall content. Often the legends themselves are cryptic and uninformative to the uninformed, {\em e.g.} ``My favorite clip''.

In order to be able to organize, categorize, summarize and index these recordings such that they can be retrieved through meaningful queries, one requires analysis of their {\em content}. Given the rather spotty nature of the metadata, the description of the content must usually be automatically derived. This naturally requires automatic identification of the {\em objects} and {\em events} that occur in the recording.  Multimedia recordings have both video and audio components. Often, the sounds in the recordings carry information that the video itself may not. Thus, not only the {\em visual} objects in the recordings be automatically detected, it is also important to detect the sounds that occur in them. 

Automatic sound event detection also finds application in other scenarios, such as monitoring traffic for sounds of accidents or impact, surveillance, where one may ``listen'' for sounds of gunshots \cite{1}, screams \cite{2} etc., which might indicate unusual noteworthy activity. It is also useful in cases such as wildlife monitoring \cite{briggs2012acoustic}, context recognition \cite{eronen} and several health and life style monitoring system.  

In all cases, the detectors themselves must be ``trained'' from examples of the sound to be detected. In general for learning such detectors, one requires annotated data, where the segments of audio containing the desired event are clearly indicated (as well as data in which the events are distinctly {\em not} present). We will refer to this type of labeling as strongly labeled data. This is fundamentally limiting, since such well-annotated data are generally scarce. 

A solution to the scarcity is to use the consumer-produced videos themselves to train the detectors.
This immediately raises several challenges, however.  Firstly, of course, the recording conditions, styles, and sophistication vary greatly among such recordings, resulting in large within-category variations between different instances of events, making the fundamental learning problem challenging. Much more important however, is the nature of annotations, if any, that they may carry. As mentioned earlier, the vast majority of consumer-produced videos carry little or no content annotation (which is what necessitates the development of automated concept detectors in the first place). Nevertheless, a significant number of these recordings {\em do} carry some weak information about their content, in the form of title, tags, etc.  By ``weak'' annotation, we mean that while they may provide information about the presence or absence of particular events in the video, they will not provide additional details such as the number of times these events occur, the precise times in the recording where they occur, or the duration of the events. These additional details are required to localize the events in the recordings in order to train detectors using conventional methods. Weak labels can be automatically inferred from metadata (tags, titles etc.) associated with recording. Hence, these weak labels must be used to train the detectors.

Conventional methods for learning classifiers for event detection generally assume the availability of datasets that contain sufficient information to isolate the segments of the recordings where the target class occurs in order to learn their characteristics. For the other kind of datasets for which only weak labels are present (like those available on internet) most methods rely on manually annotating the available data with the boundaries of events, to facilitate learning event detectors. But creating such annotations is a time-consuming and expensive procedure. Supervised learning hence becomes a difficult task.  

In this paper we take the view that it must be possible to learn event detectors from the weakly labeled data itself without requiring the additional effort of detailed annotation. Recordings that have been tagged as containing the event will have regions which are consistent with one another because they contain the event, but are not consistent with any region of recordings that are not similarly tagged. Event-carrying segments can be identified and event detectors trained by keying in on this consistency.

We embody this principle into an algorithmic framework which falls under the general rubric of {\em multiple-instance learning}. Multiple-instance learning is a generalized form of supervised learning in which labels for individual instances are not known; instead labels are available for a collection of instances, or ``bag'' as it is usually called within this setup. Although multiple-instance learning approaches have previously been employed for learning event detectors from weakly labeled data, most prior work in this area has focused on visual event recognition. The problem of learning to detect {\em acoustic} events from weakly labeled data has received very little attention. 

We propose a successful framework for acoustic event detection based on the multiple instance learning methodology. In our setup, an audio recording is considered as a bag and segments of the recording are considered as instances within the bag. Besides achieving the goal of learning acoustic event detectors using weakly labeled data; as an added benefit, we are also able to assign temporal locations to occurrences of events in the test data.

The rest of the paper is arranged as follows:  in Section 2 we give an overview of related work, in Section 3 we formulate the problem and in Section 4 we describe the proposed approach. In Section 5 we describe our experimental setup and results and we conclude in Section 6.
\section{RELATED WORK}
\label{sec:rltwrk}
The problem of multimedia content analysis has received significant attention in the past decade. Although much of the reported work relates to image- or video-based information extraction, audio-based information extraction has also been explored. These audio-based information extraction methods in general involve detection of different kinds of acoustic events such as clapping, cheering, gunshots {\em etc.} in audio components of multimedia recordings. A nice survey on  audio event detection challenges, related work, and the state of art can be found in \cite{tmmsurvey}. A brief summary of some audio event detection work is provided here. 

Possibly the most popular application so far of sound-event detection has been for surveillance \cite{1} \cite{2} \cite{tmmavsurve}. Audio-event detection has also found its way into consumer devices, particularly for automatic indexing of multimedia recordings of games \cite{huang2010}. The wider setting of detecting sound events in more generic real-world recordings has been restricted in terms of the ``vocabulary'' of sound events considered. This is mainly due to the absence of large-vocabulary open source datasets meant for audio event detection research. The authors of \cite{zhuang2010} model various sound events with Gaussian-mixture models in order to detect them in real-world recordings. The approach followed in \cite{zhuang2010} is similar to the GMM-HMM architecture used for automatic speech recognition. A simple yet effect approach converts signals into ``bags of audio words'' -- population-count histograms -- by clustering feature vectors derived from the signal to a known codebook. Classification is now performed using these bags of words as feature vectors. This approach too has been successfully applied to the problems of detecting events in audio \cite{7} as well as for multimodal approaches to event detection \cite{9} and \cite{tmmbow}. It is a general framework for obtaining a fixed length representations for audio clips and can be done on a variety of low-level audio features such as MFCCs \cite{7}, autoencoder based features \cite{autoenco} and normalized spectral features \cite{specex} to name a few. An alternate approach to obtaining bags of words is used in \cite{12} -- sound recordings are first decomposed into sequence of basic sound units called ``Acoustic Unit Descriptors'' (AUDS), which are themselves learned in an unsupervised manner. Bags of words are then obtained as bags of AUDs. The actual classification may be performed through classifiers such as SVM or random forests \cite{12}. Deep neural network based approaches have also been proposed for audio event detection \cite{gencoglu}\cite{kons} \cite{Ashraf2015}.

In every one of these cases, the detectors and classifiers that are used to detect the sound events are trained in a supervised manner. Each classifier/detector is trained using several clearly-demarcated instances of the type of sound event it must detect, in addition to several ``negative'' instances -- instances of audio segments that do not contain the event.  Obtaining data of these kinds clearly requires human annotation, and can be a time-consuming and expensive process. 

Our focus in this paper is to learn detectors from {\em weak} annotations -- annotations which only indicate the {\em presence} of sound events in recordings, without additional detail about the number of instances, their location in the recording, etc. Such annotations are, in general, easily available and easy to generate if not available. 

In comparison to supervised techniques, there has been relatively little work on learning to detect events in multimedia using weakly labeled data. Even here, the majority of published work focuses on detecting events based on {\em visual} content of the data \cite{17} \cite{tmmgeneric} \cite{18}. In the case of {\em audio}, however, literature on learning generic sound event detectors from weakly labeled data is almost negligible. Some audio-related works include music genre classification using semi-supervised approach \cite{mandel} \cite{tmmmusic}. Two other audio related works are \cite{ruizmultiple} and \cite{briggs2012acoustic} where the authors try to exploit weak labels in bird-song and bird-species classification.

In our work we perform audio event detection (AED) using weakly-labeled data by formulating it as a  {\em multiple instance learning} (MIL) problem. MIL is a learning methodology which relies on the labels of an {\em ensemble} of instances, rather than labels of individual instances. Several well known learning methods such as Support Vector Machines (SVMs)\cite{27}, $K$-NNs \cite{Wang:2000} have been modified to learn in this paradigm.
\vspace{-0.05in}
\section{AED using Weakly Labeled Data}
The problem of audio event {\em detection} is that of detecting the occurrences of events (\emph{e.g clapping, barking or cheering }) in a given audio recording.  In order to be able to do so, we require {\em models} for these events which can be used to detect their occurrence.  Training such models, in turn, requires training instances of these sounds. Such instances may be presented either as explicit recordings of these sounds, or as segments from longer audio recordings within which they occur. Our objective is to train the models, instead, using {\em weakly-labeled data} which comprise  recordings in which only the presence or absence of these events is identified, without indicating the exact location of the event or even the number of times it has occurred.

Let $R=\{R_{i}: i=1\,\,to\,\,N_R\}$ be the collection of audio recordings and $E=\{E_i:i=1\,\,to\,\,N_E\}$ be the set of events for which detection models must be built using $R$. 

For each $R_i$ a certain subset of events from $E$ are known to be present (weak-labels). For example, the information might say that $R_i$ contains events $E_1$, $E_3$ and $E_6$. It is also possible that the subset of $E$ present in $R_i$ is empty, meaning that no event from $E$ is present in $R_i$. 

Clearly, to train a detector for an event $E_i$ one cannot simply use all recordings that are marked as containing $E_i$, since a significant portion of the marked recordings might also contain other events. Moreover, since the start and end times of the occurrences of event $E_i$ in the recordings are not known, it is impossible to extract out the specific segment of the audio that contains the event for further use in supervised learning. Conventional supervised learning is thus not possible. How do we then use these weak labels for learning detectors for the events? We answer the question in the next Section. 
\vspace{-0.10in}
\section{Proposed Framework}
Our formulation of event detection using weak labels of the kind described above is based on {\em Multiple-Instance Learning} \cite{24} which is a generalized version of supervised learning in which labels are available for a collection of instances. We propose that audio event detection using weak labels is essentially an MIL problem and in general any suitable MIL algorithm can be used. 
\vspace{-0.05in}
\subsection{Multiple Instance Learning}\label{ssec:mil}
The term Multiple Instance Learning was first developed by Dietterich \emph{et al.} in 1997 for drug activity detection \cite{24}. MIL is described in terms of \textit{bags}; a \textit{bag} is simply a collection of instances. Labels are attached to the bags, rather than to the individual instances within them. A \emph{positive} bag is one which has at least one positive instance (an instance from the target class to be classified). A \textit{negative} bag contains negative instances only. A negative bag is thus pure whereas a positive bag is impure. This generates an asymmetry from a learning perspective as all instances in a \emph{negative bag} can be uniquely assigned a negative label whereas for a \emph{positive bag} this cannot be done; an instance in a \emph{positive bag} may either be positive or negative. 

Thus, it is the \emph{bag-label} pairs, rather than instance-label pairs, which form the training data from which a classifier which classifies individual instances must be learned. 

We represent the bag-label pairs as $(B_i,Y_i)$. Here $B_{i}$ is the $i^{\rm th}$ bag and contains instances $x_{ij}$ where $j=1\,\,to\,\,n_i$ and $n_i$ is the total number of instances in the bag, {\em i.e.} $B_i=\{x_{ij}\,:\,j\, = \, 1 \cdots \,n_i\}$.
$Y_i \in \{-1,1\}$ is the label for bag $B_i$. $Y_i=1$ implies a positive bag and $Y_i=-1$ implies a negative bag. We might alternatively represent negative labels by $0$ at some place for convenience (in Section \ref{ssec:milnn}). Let the total number of bags be indicated by $N$. One can attempt to infer labels of individual instances $x_{ij}$ in the bag from the bag label. The label ${y_{ij}}$ for instances in bag $B_{i}$ can be stated as:
\begin{align}
\label{eq:bag}
Y_i = -1 \Rightarrow y_{ij} = -1 \hspace{5pt}\forall \hspace{5pt} x_{ij} \in B_{i},\\
Y_i=1 \Rightarrow y_{ij} = 1 \hspace{5pt}{\rm for} \hspace{5pt} {\rm at~least~one}~ x_{ij} \in B_{i}
\end{align}
This relation between $Y_i$ and $y_{ij}$ is simply 
\begin{equation}\label{eq:baglabel}
Y_i=\max_{j}\{y_{ij}\}.
\end{equation}
The problem now is to learn a classification model $C$ so that given a new bag $\hat{B}_n$ it can predict the label $\hat{Y}_n$ for $\hat{B}_n$. Several methods have been proposed to solve the MIL problem such as Learning Axis-Parallel Concepts \cite{24}, Diverse Density\cite{25}, Citation-KNN\cite{Wang:2000}, mi-SVM and MI-SVM \cite{27}. In this work we work with two frameworks: (1) mi-SVM which modifies support vector machines for the MIL setting and, (2) neural networks for solving MIL problems
\vspace{-0.05in}
\subsection{MIL for SVM (mi-SVM)} 
\label{ssec:milsvm}
In \cite{27} Andrews \emph{et al.} proposed two methods for multiple-instance learning of Support Vector Machines. The first, called {\em mi-SVM}, operates at the instance level and maximizes the margin of individual instances from a linear discriminant. The second, referred to as {\em MI-SVM} maximizes the margin of bags of instances, rather than individual instances. In this work we use mi-SVM. 

To understand mi-SVM, we first note that the relation between bag label $Y_i$ and instance labels $y_{ij}$ can also be represented in the form of linear constraints. 
\begin{equation*}
\sum\limits_{j=1}^{n_{i}} \frac{y_{ij}+1}{2} \ge 1\,\,\forall \,\,i\,\, s.t\,\, Y_i=1,\,\,; \,\, y_{ij}=-1 \,\forall \,\,i\,\, s.t\,\, Y_i=-1
\vspace{-0.1in}
\end{equation*}
The instance labels in mi-SVM are treated as unobserved integer variables subject to the constraints defined above. As in conventional training of SVMs, where we must estimate the parameters of a linear (or kernelized) discriminant function such that the {\em margin} of training instances from the discriminant function is maximized,  here too we must maximize the margin. However, since the instance labels are unknown, we modify the objective: the goal now is to maximize a {\em soft margin} over both the decision function and the hidden integer variables, namely the unknown labels of instances. Optimization of the generalized soft margin is thus
\begin{equation} \label{eq:mar}
\min_{y_{ij},\textbf{w},b,\xi} \frac{1}{2}||w||^2 + C\sum\limits_{ij}\xi_{ij}
\end{equation}
such that 
\begin{align}
\forall i,j\,:\, y_{ij} (\langle \textbf{w} ,\textbf{x}_{ij} \rangle + b ) \, \ge 1-\xi_{ij}\\
\xi_{ij}\,\ge 0 \,\,; y_{ij}\,\in\,\{-1,1\} \,\,; y_{ij}=-1 \,\forall \,\,i\,\, s.t\,\, Y_i=-1\\
\sum\limits_{j=1}^{n_{i}} \frac{y_{ij}+1}{2} \ge 1\,\,\forall \,\,i\,\, s.t\,\, Y_i=1
\end{align}
In equation \ref{eq:mar} the labels  $y_{ij}$ of instances belonging to positive bags are unknown integer variables. As a result both, the optimal labeling of these instances as well as the optimal hyperplane $(\textbf{w}, b)$, must be computed. The separating hyperplane must be such that there is at least one pattern from every positive bag in the positive half space and all patterns belonging to negative bags are in the negative half space.

The above formulation is, however, a difficult mixed integer problem to solve. Andrews {\em et al.} \cite{27} proposed an optimization heuristic to solve this integer problem. The main idea behind the heuristic is that for given integer variables {\em i.e.} fixed labels, it can be solved exactly through usual quadratic programming. The solution thus is a two step iterative process:
\begin{itemize}
\item \textit{Step 1:} Given the integer variables (fixed labels), solve the standard SVM. 
\item \textit{Step 2:} Given the SVM solution, impute the integer label variables for the positive bags.
\end{itemize}
The labels of instances in the positive bags are all initialized as positive and are updated as described above. The instances in negative bags are obviously labeled negative and remain so through out the procedure. The two steps are iterated until no changes in labels occur. The overall mi-SVM is shown in Algorithm \ref{miSVM}.
\begin{algorithm}[t]
\caption{mi-SVM Algorithm}\label{miSVM}
\begin{algorithmic}[1]
\Procedure{Learning SVM in MIL setting}{$B_i,Y_i$} // Input training bags and labels\newline\newline
//Initialization Step 
\State $y_{ij}=Y_{i}$ for all j in Bag $B_{i}$
\Repeat
\State {\small compute SVM solution $\textbf{w}$, $b$ with imputed labels}
\For{all bag $B_{i}$ s.t $Y_{i}=1$}
\State {\small compute $f_{ij}=(\langle \textbf{w} ,\textbf{x}_{ij} \rangle + b )$ $\forall$ $x_{ij}$ in $B_i$}
\State $y_{ij}=sgn(f_{ij})$ $\forall$ $j$ in bag $B_{i}$ s.t $Y_{i}=1$
\If{$(\sum\limits_{j=1}^{n_{i}} \frac{y_{ij}+1}{2}==0)$} 
\State compute $j^{*}=\arg\max_{j}(f_{ij})$
\State set $y_{ij^{*}}$=1
\EndIf
\EndFor
\Until imputed labels no longer change
\EndProcedure
\end{algorithmic}
\end{algorithm}
If it happens during an iteration that all instances in a positive bag are labeled as negative, then the one with the maximum value for discriminant function is assigned a positive label. This process is repeated until no change in imputed labels is observed.

\subsection{MIL of Neural Networks (BP-MIL)} \label{ssec:milnn}
Neural networks have become increasingly popular for classification tasks \cite{schmidhuber} \cite{lecun}. The conventional approach to training neural networks is to provide instance-specific labels for a collection of training instances. Training is performed by updating network weights to minimize the average divergence between the actual network output in response to these training instances and a desired output, typically some representation of their assigned labels \cite{rumelhart}\cite{werbos1981}. 

In the MIL setting, where only bag-level labels are provided for the training data, this procedure must be appropriately modified. In order to do so, we must modify the manner in which the divergence to be minimized is computed to utilize only bag-level labels. For this, we employ an adaptation of neural networks for multiple instance learning (BP-MIL) proposed in \cite{28}.  

Let $o_{ij}$ represent the output of the network in response to input $x_{ij}$, the $j^{\rm th}$ instance in $B_i$, the $i^{\rm th}$ bag of training instances. 
We define the {\em bag-level} divergence for bag $B_i$ as
\begin{equation}
E_i=\frac{1}{2}\left(\max_{1 \le j \le n_i}(o_{ij})-d_i\right)^2
\label{eq:divx}
\end{equation}
where $d_i$, the desired output of the network in response to the set of instances from $B_i$, is simply set to $Y_i$, the label assigned to $B_i$. Thus for positive bags $d_i = 1$, whereas for negative bags $d_i = 0$.
The central idea behind the bag-level divergence of Equation \ref{eq:divx} is to refer to any bag using the instance which produces the maximal output. This was proposed by Dooly \emph{et al.} in \cite{29} where they showed that irrespective of the number of instances (positive or negative) in a bag,  the bag can be fully described by the instance with maximal output.

The bag-level divergence of Equation \ref{eq:divx} may be understood by noting that the term ``$\max_j o_{ij}$'' in it effectively represents the {\em bag-level output} of the network, and that Equation \ref{eq:divx} simply computes the divergence of the bag-level output with respect to the bag-level label of Equation \ref{eq:baglabel}. 

The ideal output of the network in response to any negative instance is $0$, whereas for a positive instance it is $1$. For {\em negative} bags, Equation \ref{eq:divx} characterizes the {\em worst-case} divergence of all instances in the bag from this ideal output. Minimizing this effectively ensures that the response of the network to {\em all} instances from the bag is forced towards $0$. In the ideal case, the system will output $0$ in response to all inputs in the bag, and the divergence $E_i$ will go to $0$.

For {\em positive} bags, on the other hand, Equation \ref{eq:divx} computes the {\em best-case} divergence of the instances of the bag from the ideal output of $1$. Minimizing this ensures that the response of the network to {\em at least} one of the instances from the bag is forced towards $1$. In the ideal case, one or more of the inputs in the bag will produce an output of $1$ and the divergence $E_i$ will go to zero. 

The {\em overall} divergence on the training set is obtained by summing the divergences of all the bags in the set: 
\begin{equation} \label{eq:oerr}
E=\sum\limits_{i=1}^N E_i =\sum\limits_{i=1}^N \frac{1}{2}\left(\max_{1 \le j \le n_i} o_{ij}-d_i\right)^2
\end{equation}
The parameters of the network are trained using conventional backpropagation, with the difference that we now compute gradients of the divergence given in Equation \ref{eq:oerr}, and that entire bags of data must be processed prior to updating network parameters. During training once all instances in a bag have been fed forward through the network, the weight update for the bag is done with respect to the instance in the bag for which the output was maximum. The process is continued until the overall divergence falls below a desired tolerance. 

Prediction using a trained network can now be done instance-wise as in done in classical feed-forward neural networks. Bag labels, if required, can be predicted based on the label obtained for the maximal-scoring instance in the bag. 
\vspace{-0.05in}
\subsection{MIL for AED using weakly labeled data} 
\label{ssec:milaudio}
In our setting for training audio-event detectors from weakly labeled data, we only have labels informing us whether a recording contains a given event or not. In order to apply the MIL framework to this scenario, we must first represent the weakly labeled recordings in $R$ in terms of bag-label representations. 

To convert recording $R_{i}$ to a bag, it is segmented into a number of short audio segments. Adjacent segments may overlap by design. Let the segments derived from $R_i$ be $[I_{R_{i}1}\,I_{R_{i}2}\,...I_{R_{i}K}]$.  Each of these smaller segments is now treated as an individual instance within the bag $R_{i}$. Formally a bag $R_i$ is thus $R_i=\{I_{R_{i}1},I_{R_{i}2},....I_{R_{i}K}\}$
where $K$ depends upon the duration of recording $R_i$, the length of the individual segments and the overlap between adjacent segments.  

If the weak labels for $R_i$ mark an event as being present in $R_i$, then it will be present in at least one of the instances within it. Hence, $R_i$ will be a positive bag for the instance. On the other hand, if an event is marked as not being present in a recording, then clearly  \emph{none} of the segments(instances) from the recording will be positive for that event, and hence overall that recording is a negative bag for the event. Hence, the weak labels for all the recordings can directly provide bag-label representations needed in MIL. The MIL approaches can now be used to learn a model which can predict the presence or absence of an event in each recording, and even identify the locations of the events within the recordings. 
\vspace{-0.05in}
\subsection{Temporal Localization of Events} 
\label{ssec:miltle}
The MIL frameworks we use in this work learn from bag-level labels; but once learning is complete, they can classify individual instances. In the context of audio analysis, this implies that not only can we detect the presence of an event in a test recording (bag) but also in individual segments of the test recording. Formally, if $R_x$ is a test recording, which we may more explicitly represent as $R_x(t)$ where ``$t$'' indexes time,  the individual segments in the recording, $I_{R_{x}1},I_{R_{x}2},....I_{R_{x}K}$, are given by $I_{R_{x}k} = R_x(t),~~(k-1)l^{'} \leq t < (k-1)l^{'}+l $, where $l$ is the length of the segment in seconds and $l^{'}$ denotes the amount by which a segment window is shifted with respect to the previous segment. In the special case of non-overlapping segments $l^{'}=l$. If an event $E_i$ is detected in segment $I_{R_{x}k}$ it means this event can be localized to the time segment $\left((k-1)l^{'},~(k-1)*l^{'}+l\right)$ in the recording $R_x$. Hence, the framework can generate information about the temporal location of events. Hence, we are able to obtain a complete description of the recording in terms of audio events. This is a unique form of AED learning since the descriptive form of labeling was never present in the training data in the first place. 
\subsection{GMM based features for audio segments}
\label{ssec:milgmm}
Before we can apply the MIL framework, each segment of audio must first be converted to an appropriate feature representation. We use Mel-frequency cepstral coefficient (MFCC) vectors to obtain low-level feature representations for audio segments. However, direct characterization of audio as sequences of MFCC vectors tends to be ineffective for the purposes of audio classification \cite{zhuang2010}; other secondary representations derived from these are required. As described in Section \ref{sec:rltwrk}, one simple and yet very successful approach is the Bag of Audio Words feature representation, which quantizes the individual MFCC vectors into a set of codewords, and represents segments of audio as histograms over these codewords.

However, while the bag of words representation has been found to be very effective for classification of long segments of audio, for shorter duration segments such as those in our case, they present several problems. Bag of words representations effectively characterize the {\em distribution} of the MFCC vectors in the segment. However, because of the inherent quantization they lose much of the detail of this distribution, which is required for fine-level analysis, such as in the detection of sound events, particularly when the classifiers must be learned from weak labels. While the loss of resolution may be partially resolved by increasing the size of the codebook used for quantization, this can lead to generation of sparse histograms for short audio segments, with large cross-instance variation in the derived features.

In \cite{23} it was demonstrated that for short audio segments characterization of the audio using a Gaussian Mixture Model (GMM) can provide robust representations for  performance of audio event detection. They suggested Gaussian Mixture based characterization of audio events is a combination of two features: the first, which we represent as $\vec{F}$, is similar to bag-of-words characterizations such as \cite{van2010}, and the second, which we represent as $\vec{M}$ is a characterization of the modes of the distribution of vectors in the segment. 

As a first step to obtaining the $\vec{F}$ and $\vec{M}$ feature vectors for the audio segments, we train a {\em universal} Gaussian mixture model (GMM) on MFCC vectors from a large and diverse collection of audio recordings. This background GMM is used to extract the $\vec{F}$ and $\vec{M}$ features.  In the following, we will represent this universal GMM as $\mathcal{G} = \{w_k,N(x; \lambda_k)\}$, where $w_k$ is the {\em a priori} probability or  mixture weight of the $k^{\rm th}$ Gaussian in the mixture, $N(.)$ represents a Gaussian, and $\lambda_k$ collectively represents the set of mean and covariance parameters of the $k^{\rm th}$ Gaussian.
\vspace{-0.05in}
\subsubsection{$\vec{F}$ Features}
\vspace{-0.05in}
For each audio segment we have a sequence of $D$-dimensional MFCCs vectors denoted by $\vec{x_{t}}$ where $t=1\,\,to\,\,T$. $T$ is the total number of MFCC vectors for the given segment. For each component $k$ of the background GMM we compute
\begin{align}
Pr(k | \vec{x_{t}})=& \frac{w_{k}N(\vec{x_{t}} ; \lambda_k)}{\sum\limits_{j=1}^G w_jN(\vec{x_{t}} ; \lambda_j)}, \\
F(k) =& \frac{1}{T}\sum\limits_{i=1}^T Pr(k | \vec{x_{t}})
\end{align}
The $\vec{F}$ feature vectors are $\vec{F} = [F(1), F(2), \cdots, F(G)]^\top$. Thus, $\vec{F}$ is $G$-dimensional vector representing a normalized \textit{soft-count} histogram of the MFCC vectors in the recording. It captures how the MFCC vectors are distributed across components of $\mathcal{G}$. It is a variant of bag of audio word features where soft assignment is used in place of hard quantization. 
\vspace{-0.05in}
\subsubsection{$\vec{M}$ Features}
A more detailed characterization can be obtained by actually representing the distribution of the feature vectors in the segment. To do so, we train a separate GMM for each audio segment by adapting the universal GMM $\mathcal{G}$ to the collection of MFCC vectors in the segment. The means of the universal GMM are adapted to each training segment using the maximum {\em a posteriori} (MAP) criterion as described in \cite{bimbot}. This is done as follows for $k^{th}$ component of the mixture
\begin{align}
n_{k}=&\sum\limits_{t=1}^T Pr(k | \vec{x_{t}}), \\
E_{k}(\vec{x})=&\frac{1}{n_{k}}\sum\limits_{t=1}^T Pr(k | \vec{x_{t}})\vec{x_{t}}
\end{align}
Finally the updated means are computed as 
\begin{equation}
\hat{\vec{\mu_k}}=\frac{n_k}{n_k+r}E_{k}(\vec{x})+\frac{r}{n_k+r}\vec{\mu_k}
\end{equation}
where $\vec{\mu_k}$ is the mean vector of $k^{th}$ Gaussian and $r$ is a relevance factor. The means of all components are then appended to form the $G \times D$ vector $\vec{M}$ as  $\vec{M} = [\hat{\vec{\mu_1}}^\top, \hat{\vec{\mu_2}}^\top,\cdots \hat{\vec{\mu_G}}^\top]^\top$. 
The above two features are unsupervised methods of characterizing the statistical structure of an audio segment. $\vec{F}$ is the coarser representation, but can however be robustly estimated. $\vec{M}$ is more detailed, though it is also more easily affected by inter-instance variability. Together, they give us a robust representation of both the coarse and fine structure of the signals in short audio segments. In our experiments we have used $\vec{F}$ both in combination with $\vec{M}$, and as a standalone feature.

Once we have a robust set of features for representing events in short audio segments we can extract features for each audio segment (instances) of a recording (bag). Thus a recording(bag) $R_i=\{I_{R_{i}1},I_{R_{i}2},....I_{R_{i}K}\}$ in feature space becomes $ R_i=\{\vec{x}_{{R_{i}1}};\vec{x}_{{R_{i}2}};....\vec{x}_{{R_{i}K}}\} $ where $\vec{x}_{{R_{i}j}}$ are either the $\vec{F}$ features alone or the concatenated $\vec{F}$ and $\vec{M}$ vectors. The bags are then fed into the the MIL frameworks BP-MIL or mi-SVM for learning event detector models.
\section{Experiments and Results}
\label{sec:pagestyle}
We evaluated the proposed MIL framework on a portion of the TRECVID-MED 2011 database \cite{32}. The videos in this dataset are meant for multimedia event detection and belong to broad categories such ``Changing a vehicle tire'', ``Attempting a board trick'', ``feeding an animal'', {\em etc} which are not particularly suitable for audio event detection study. Hence we work with more meaningful acoustic events such as \emph{clapping, cheering} etc. A subset of the MED dataset is thus annotated with 10 such events. To be able to compare performance with the fully supervised case and to compute performance metrics for temporal  localization of events, our annotations include actual locations and duration of occurrences of these events. However, only the information regarding the presence or absence of these sounds is used in our MIL based framework. 

A total 457 recordings (bags) are used in the experiments. This is over 22 hours of audio data. We henceforth refer to this set of recordings as the ``dataset''. The length of each recording in the dataset varies from a few seconds to several minutes with an average length of about 2.9 minutes. This implies that the number of instances in each bag also has a wide range. 

Ideally, the length of each segment should be properly set keeping in mind the expected duration of event. However, we observed that segment length decided by heuristics work well. Since the median length of the chosen sound events was less than 1 second, we will report results for all our results on segment length fixed to $1$ seconds. The segments overlap by $50\%$. This results in well over 150,000 total instances. The names of the 10 events and the total number of positive bags for each events are given in Table \ref{tab:bagsz}. It is worth noting that some of the recordings in the dataset do not contain any of the 10 events; also a recording might be a positive bag for more than 1 event. Hence, the sum total of numbers in Table \ref{tab:bagsz} is different from the total number of recordings in the dataset.
\begin{table}[t]
\centering
\caption{Number of Positive Bags for each event }
\begin{tabular}{|c|c|}
\hline  
Events & Number of Bags\\
\hline 
Cheering & 171 \\
\hline
Children's Voices & 33 \\
\hline
Clanking & 13\\
\hline 
Clapping & 102 \\
\hline
Drums & 25 \\
\hline
Engine Noise & 80 \\
\hline
Hammering & 17 \\
\hline
Laughing & 116 \\
\hline
Marching Band & 24\\
\hline
Scraping & 30\\
\hline
\end{tabular}
\label{tab:bagsz}
\vspace{-0.15in}
\end{table}
As is clear from our proposed framework our goal is detection of audio events in recordings. Hence, the training data for the binary classifiers  for each event have positive bags equal to the number shown in Table \ref{tab:bagsz}, while the rest of the recordings in the dataset are negative bags for that event.  The dataset is partitioned into 4 sets. Three of the sets were used to train the models, which were then tested on the fourth set. This was done in all four ways meaning each set becomes a test set. This gives us results on the whole dataset. Hence, all results reported here are on the entire dataset. 

All recordings were parameterized into sequences of 21-dimensional Mel Frequency Cepstrum Coefficient (MFCC) vectors. MFCC vectors were computed over analysis frames of 20ms, with an overlap of 50$\%$(10ms) between adjacent frames. The $\vec{F}$ and $\vec{M}$ features were derived from these sequences of MFCC vectors. We trained two background GMMs with 64 and 128 Gaussian components respectively. The number of Gaussian component in the features is represented as subscript in the feature such as $\vec{F}_{64}$, $\vec{M}_{64}$. 

ROC curves are used to analyze the performance. ``Area Under ROC curve'' (AUC) \cite{bradleyROC} is a well known metric used to characterize ROC curves. AUC is used to compare results in different cases. Higher AUC values indicate better detection performance. We first show results for detection of events at recording (bag) level using miSVM and BPMIL in Section \ref{sec:misvmres} and \ref{sec:bpmilres} respectively. The instance level results for temporal localization of events are provided in Section \ref{sec:tempres}. 
\subsection{miSVM Results}
\label{sec:misvmres}
For the miSVM framework linear SVMs are used in all experiments. We use LIBLINEAR \cite{fanliblinear} in our implementation of the miSVM framework. The slack parameter $C$ in the SVM formulation is obtained by 4 fold-cross validation over the training set. A comprehensive analysis through comparison of results in different cases is provided. The mean AUC over all events is shown in the last row of each table. 
\begin{table}[t]
\centering
\caption{AUC comparison with supervised SVM}
\begin{tabular}{|c|c|c|}
\hline  
Events & AUC ($miSVM$) & AUC ($supSVM$)\\
\hline 
Cheering & 0.632 & 0.682 \\
\hline
Children's Voices & 0.678 & 0.668 \\
\hline
Clanking & 0.714 & 0.727\\
\hline 
Clapping & 0.646 & 0.697\\
\hline
Drums & 0.60 & 0.640\\
\hline
Engine Noise & 0.623 & 0.671 \\
\hline
Hammering & 0.557 & 0.568 \\
\hline
Laughing & 0.527 & 0.741 \\
\hline
Marching Band & 0.551 & 0.558\\
\hline
Scraping & 0.723  & 0.850\\
\hline
\textbf{Mean} & \textbf{0.625} & \textbf{0.680}\\
\hline
\end{tabular}
\label{tab:supcmp}
\vspace{-0.1in}
\end{table}
\subsubsection{Comparison with supervised SVM}
We start by showing comparison of our proposed framework with fully supervised AED where strong labels are available. For supervised learning the time stamps in the annotations are used to obtain pure examples of each event, following which an SVM is trained using feature representations of these examples. Table \ref{tab:supcmp} shows this comparison. Comparison is shown for $\vec{F}_{64}$ features. In Table \ref{tab:supcmp} ``\emph{supSVM}'' refers to supervised SVM. As is expected, supervised SVMs perform better than \emph{miSVM}. However, there are several events for which the performance obtained with weak labels is comparable to that obtained with strong labels. Although the performance of supervised SVMs can potentially be improved by obtaining more strongly labeled examples for each event, Table \ref{tab:supcmp} illustrates that \emph{miSVM} too can achieve fairly decent performance using only weak labels. 
\subsubsection{Number of Gaussian Components}
Table \ref{tab:gscomp} shows AUC results for $\vec{F}_{64}$ and $\vec{F}_{128}$ features. It can be noted that there are several events for which increasing the number of Gaussians leads to about $2-4 \%$ absolute improvement in AUC values. At the same time there are events such as {\em Cheering} and {\em Marching Band} where this improvement is not observed, or the performance goes down as in {\em Children's Voices}. A performance drop of about $4\%$ is observed in this case. The optimal cluster size is known to be event specific in AED, and this holds for MIL-based audio event detection as well. 
\begin{table}[t]
\centering
\caption{AUC for different number of components in GMM (miSVM)}
\begin{tabular}{|c|c|c|}
\hline  
Events & AUC ($\vec{F}_{64}$) & AUC ($\vec{F}_{128}$)\\
\hline 
Cheering & 0.632 & 0.638 \\
\hline
Children's Voices & 0.678 & 0.633 \\
\hline
Clanking & 0.714 & 0.744\\
\hline 
Clapping & 0.646 & 0.667\\
\hline
Drums & 0.60 & 0.636\\
\hline
Engine Noise & 0.623 & 0.642 \\
\hline
Hammering & 0.557 & 0.587 \\
\hline
Laughing & 0.527 & 0.540 \\
\hline
Marching Band & 0.551 & 0.554\\
\hline
Scraping & 0.723  & 0.735\\
\hline
\textbf{Mean} & \textbf{0.625} & \textbf{0.637}\\
\hline
\end{tabular}
\label{tab:gscomp}
\vspace{-0.15in}
\end{table}
\begin{table}[t]
\centering
\caption{Effect of $\vec{M}$ features addition (miSVM)}
\begin{tabular}{|c|c|c|}
\hline  
Events & AUC ($\vec{F}_{64}$) & AUC ($[\vec{F}_{64}, \vec{M}_{64}]$)\\
\hline 
Cheering & 0.632 & 0.668 \\
\hline
Children's Voices & 0.678 & 0.723 \\
\hline
Clanking & 0.714 & 0.859\\
\hline 
Clapping & 0.646 & 0.680\\
\hline
Drums & 0.60 & 0.639\\
\hline
Engine Noise & 0.623 & 0.575 \\
\hline
Hammering & 0.557 & 0.660\\
\hline
Laughing & 0.527 & 0.641 \\
\hline
Marching Band & 0.551 & 0.745\\
\hline
Scraping & 0.723  & 0.744\\
\hline
\textbf{Mean} & \textbf{0.625} & \textbf{0.693}\\
\hline
\end{tabular}
\label{tab:mfeat}
\vspace{-0.15in}
\end{table}

\subsubsection{Adding $\vec{M}$ Features}
We now observe the effect of adding the $\vec{M}$ features to the system, along with $\vec{F}$. Table \ref{tab:mfeat} shows a comparison of AUC values when only $\vec{F}$ is used and when it is combined with $\vec{M}$ features. The results shown are obtained with $64$ Gaussian components in the GMM. It can be observed that adding $\vec{M}$ features obtained by  {\em maximum a posteriori} adaptation leads to remarkable improvement of results for almost all events. Events for which $\vec{F}$ features alone results in very poor performance such as {\em Hammering}, {\em Laughing} and {\em Marching Band} benefit significantly from the $\vec{M}$ features. Absolute improvements of $10.3\%$, $11.4\%$ and $19.4\%$ respectively are observed for these three events. For other events too absolute improvements in the range of $2.1\% - 14.5\%$ can be noted. The only exception is {\em Engine Noise} for which the \emph{soft-count} $\vec{F}$ seems to be better. It is likely that although we observe improvements in miSVM, the actually improvements may be classifier dependent.

\subsubsection{Overall Results}
\begin{figure}[t]
\centering
\includegraphics[width=0.49\linewidth,height=1.5in]{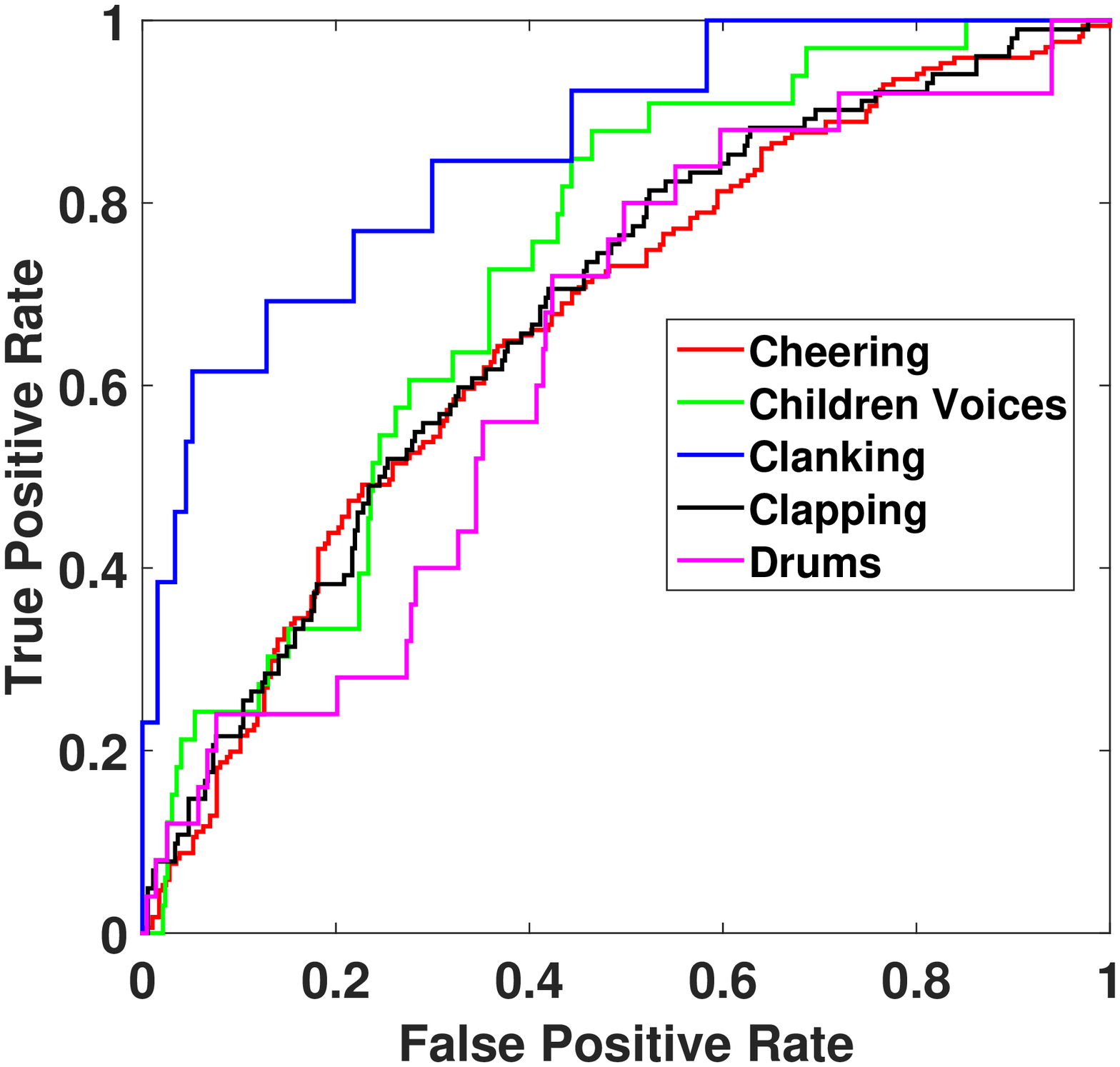}
\includegraphics[width=0.49\linewidth,height=1.5in]{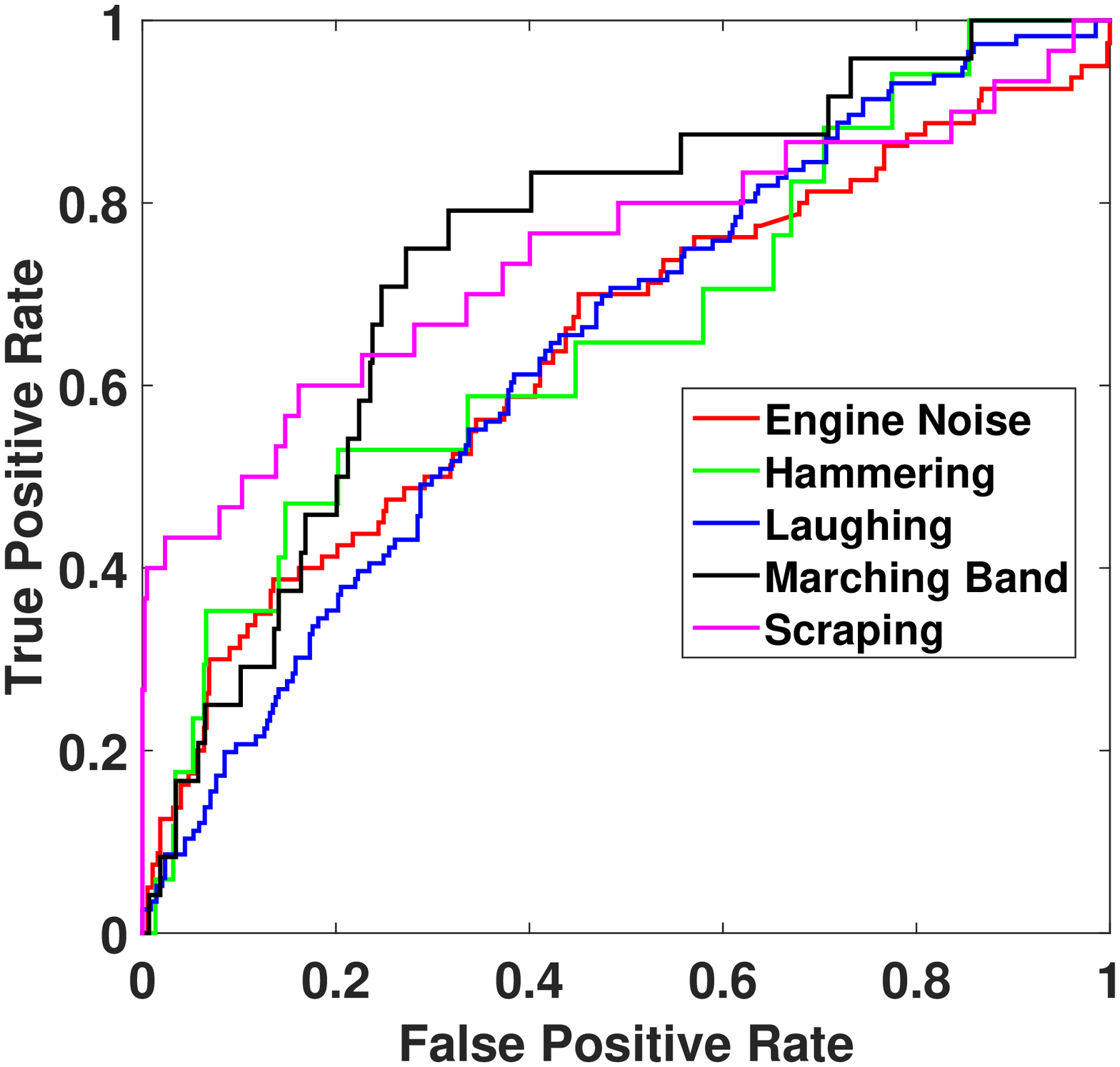}
\caption{ROC Curves for different events using miSVM framework (Best Viewed in Color)}
\label{fig:rocsvm}
\vspace{-0.15in}
\end{figure}

The overall AUC results across all experiments using miSVM is shown in Table \ref{tab:bestmisvm}. This is the best result across different feature representations for audio segments. The corresponding ROC curves are in Figure \ref{fig:rocsvm}. Events such as {\em Clanking}, {\em Children's's Voices}, {\em Scraping} and {\em Marching Band} are easier to detect compared to other events such as {\em Drums}.  The mean AUC over all events is $0.704$ which validates the success of our proposed framework.

\begin{table}[t]
\centering
\caption{Overall Results with miSVM }
\begin{tabular}{|c|c|c|c|}
\hline  
Events & AUC & Events & AUC\\
\hline 
Cheering & 0.668 & Engine Noise & 0.642\\
\hline
Children's Voices & 0.730 & Hammering & 0.660\\
\hline
Clanking & 0.859 & Laughing & 0.685 \\
\hline 
Clapping & 0.680 & Marching Band & 0.745\\
\hline
Drums & 0.639 & Scraping & 0.744 \\
\hline
\multicolumn{2}{|c|}{\textbf{Mean}} & \multicolumn{2}{c|}{\textbf{0.704}}\\
\hline
\end{tabular}
\label{tab:bestmisvm}
\vspace{-0.15in}
\end{table}
\vspace{-0.05in}
\subsection{BP-MIL Results}
\label{sec:bpmilres}
For the BP-MIL neural network three parameters must be defined, namely, number of hidden layers, the number of nodes in each hidden layer ($n_{no}$) and learning rate($\eta$). We used a network with one hidden layer in all experiments. The network is trained for a total of $60$ epochs. The learning rate is either fixed at $0.1$ throughout training or $0.1$ for the first $30$ epochs and then reduced in each epoch till it reaches $0.01$. Larger values of $n_{no}$ are used for larger dimensionality of input features. For $\vec{F}_{64}$  and $\vec{F}_{128}$ features 3 different values of $n_{no}$ are used. These are $16,50$ and $100$ for $\vec{F}_{64}$ and $50,100$ and $150$ for $\vec{F}_{128}$. When both $\vec{F}$ and $\vec{M}$ are used, the values of $n_{no}$ used in the experiments are $256$ and $512$. Training neural networks in general requires exhaustive tuning of parameters to get good results. Although results presented here show reasonable performance for the BP-MIL framework, we believe that better results can be obtained by more aggressive parameter tuning. In fact parameter tuning may give better insight into the BP-MIL framework. 

\begin{figure}[t]
\centering
\includegraphics[width=0.49\linewidth,height=1.5in]{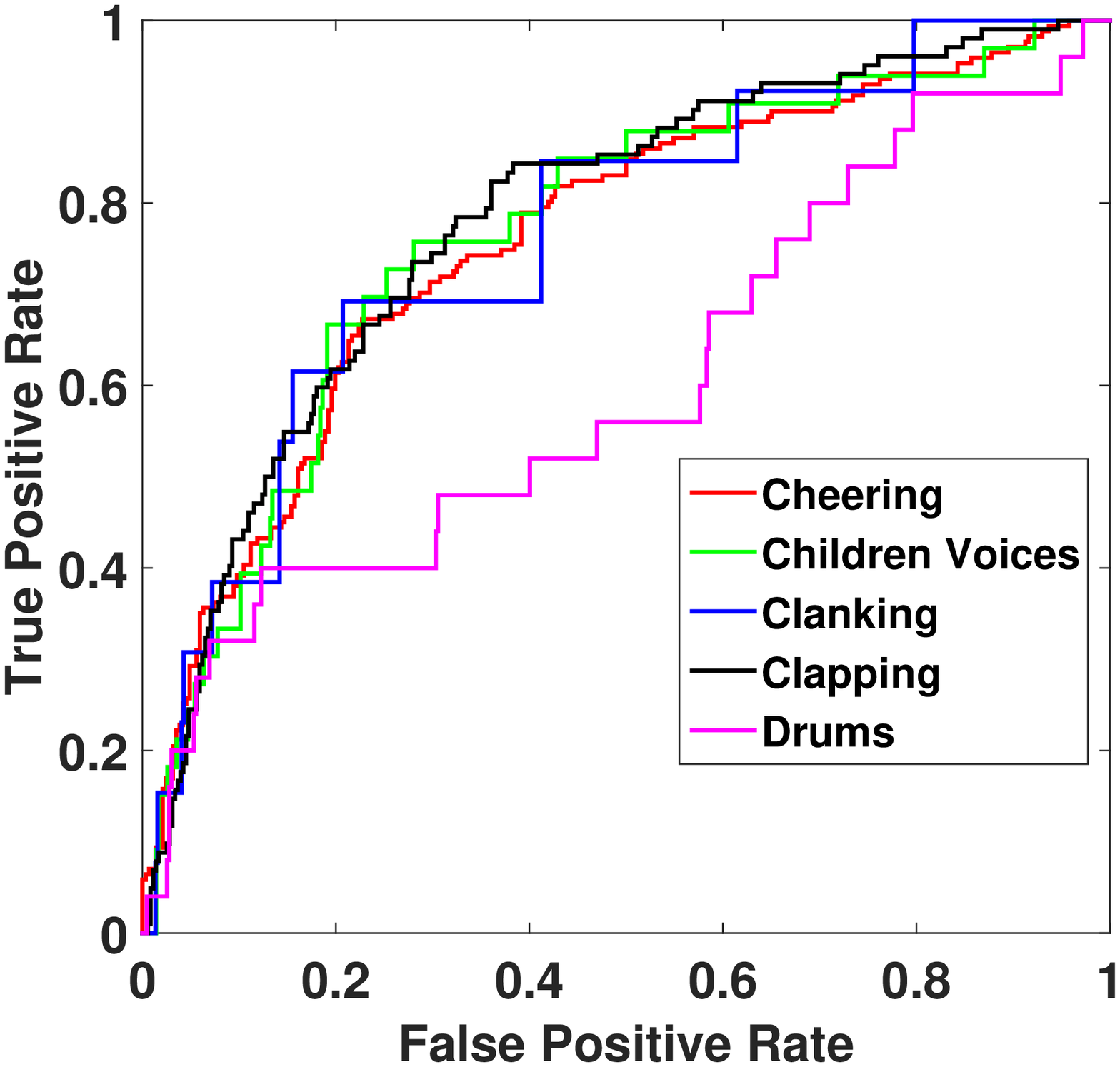}
\includegraphics[width=0.49\linewidth,height=1.5in]{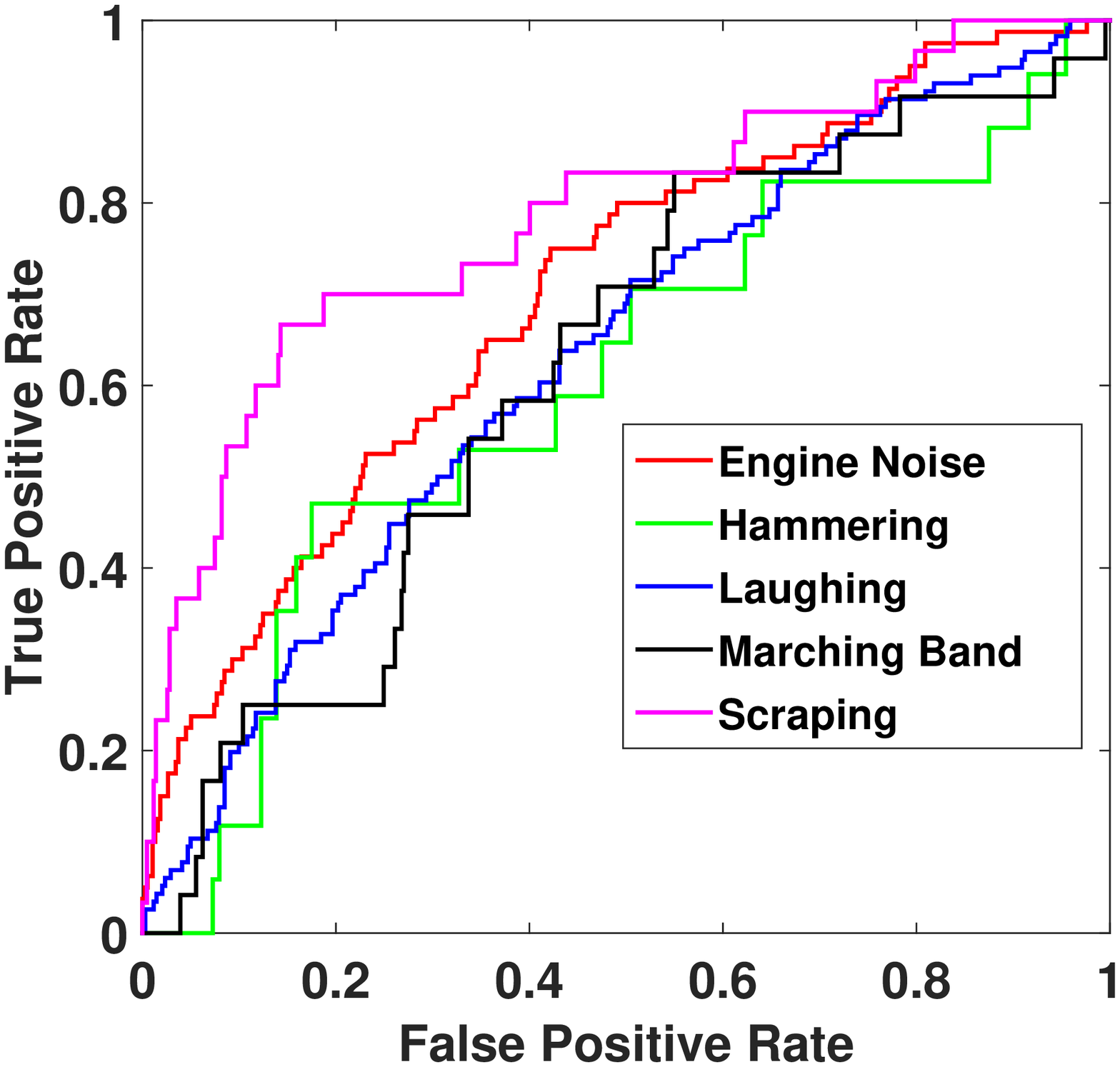}
\caption{ROC Curves for different events using BP-MIL framework (Best Viewed in Color)}
\label{fig:rocbpmil}
\vspace{-0.15in}
\end{figure}

We present only the best results across the different settings. Before looking into these results, we make note of the fact that for BP-MIL adding $\vec{M}$ features in general leads to poorer performance. Some improvement is observed only for {\em Marching Band} and {\em Drums}. One possible reason for this might be the substantial increase in total number of weight parameters with the addition of $\vec{M}$ features. As the network size increases by a considerable amount due to increase in input feature size, a substantial increase in training data is expected for learning the model. However, in the present case, the size of our data remains unchanged; this might be one of the reasons for the poor performance for BP-MIL on the addition of $M$ features. It is possible that with a larger dataset $M$ features may be beneficial in the neural network setting as well.  Standalone $\vec{F}$ work well for BP-MIL. Table \ref{tab:bestbpmil} shows overall results for BP-MIL framework. The corresponding ROC curves are shown in Figure \ref{fig:rocbpmil}. If we compare these results with miSVM approach we can observe that events such as {\em Scraping}, {\em Clanking} and {\em Children's Voices} are easier to detect in this case as well. Events such as \emph{Drums} and \emph{Hammering} are harder to detect using BP-MIL as well. The mean AUC remains almost same as miSVM, however, one can note a significant difference with respect to miSVM for several events. An analysis on a larger vocabulary of events might help differentiate the two approaches more clearly.
\begin{table}[t]
\centering
\caption{Overall Results with BP-MIL }
\begin{tabular}{|c|c|c|c|}
\hline  
Events & AUC & Events & AUC\\
\hline 
Cheering & 0.759 & Engine Noise & 0.698\\
\hline
Children's Voices & 0.767 & Hammering & 0.603\\
\hline
Clanking & 0.764 & Laughing & 0.632 \\
\hline 
Clapping & 0.781 & Marching Band & 0.618\\
\hline
Drums & 0.601 & Scraping & 0.785 \\
\hline
\multicolumn{2}{|c|}{\textbf{Mean}} & \multicolumn{2}{c|}{\textbf{0.701}}\\
\hline
\end{tabular}
\label{tab:bestbpmil}
\vspace{-0.15in}
\end{table} 
\vspace{-0.05in}
\subsection{Temporal Localization of Events}
\label{sec:tempres}
We now show the performance of the MIL based framework on temporal localization of audio events. To evaluate the performance on this task we need the ground truth labels of all instances in all bags. The instances in the bags have been obtained through uniform segmentation in our work. Each instance is a one second window segment of the recording which is moved in an overlapping manner to segment the recording. However, the annotations providing time stamps of events in the recording does not adhere to this uniform segmentation. Thus an event might start and end within a segment and it can also start or end at any point in the segment. Hence, assigning ground truth labels of instances for a valid analysis is not straightforward. We use a simple heuristic to obtain ground truth labels. As described in Section \ref{ssec:miltle} each segment represents a specific time duration of the recording. Looking into the actual annotations available, if an event can be marked to be present in at least $50\%$ of the total length of the segment we call the ground truth label of that segment as positive;  otherwise it is negative. 

\begin{figure}[t]
\centering
\includegraphics[width=0.49\linewidth,height=1.5in]{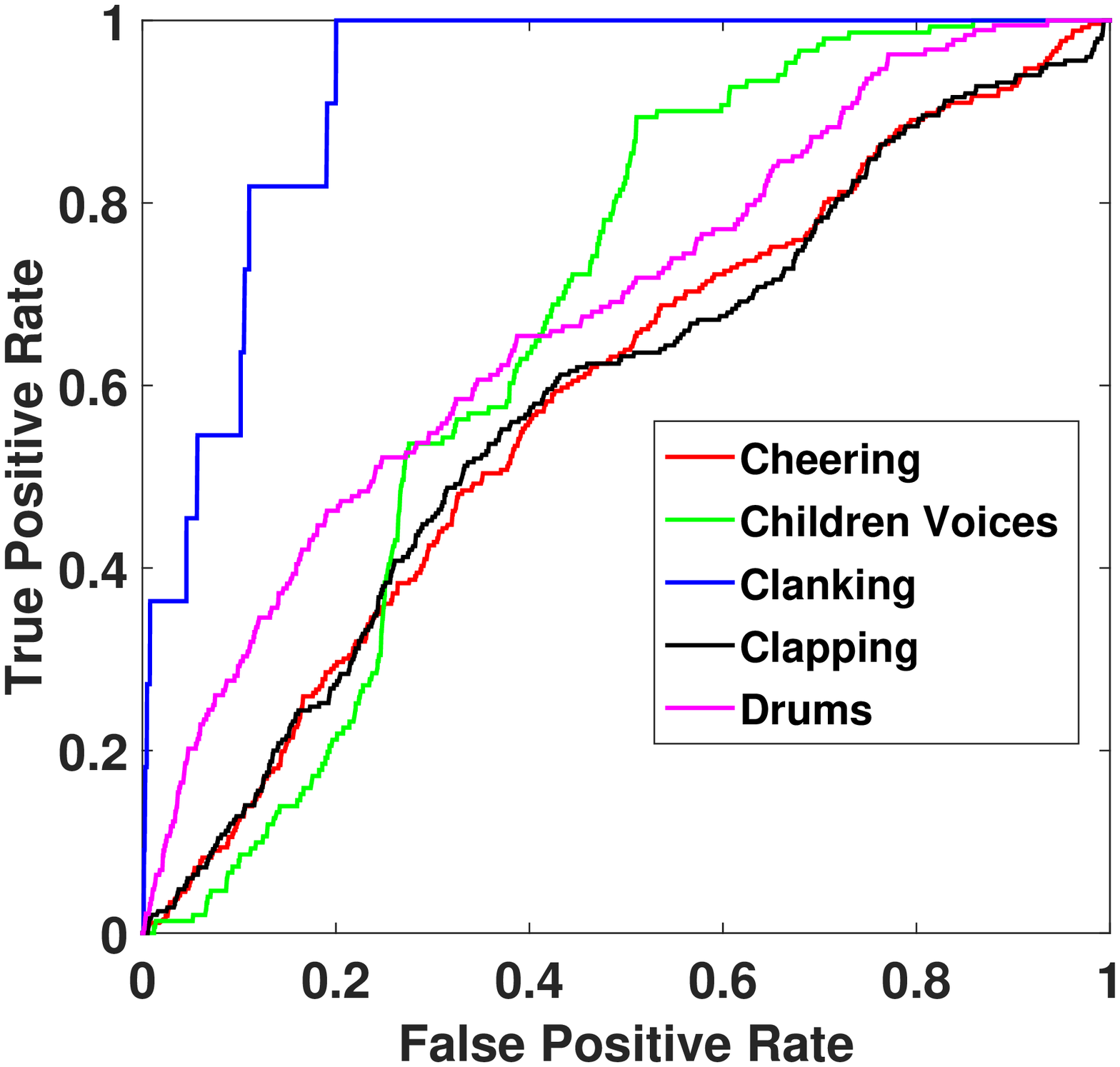}
\includegraphics[width=0.49\linewidth,height=1.5in]{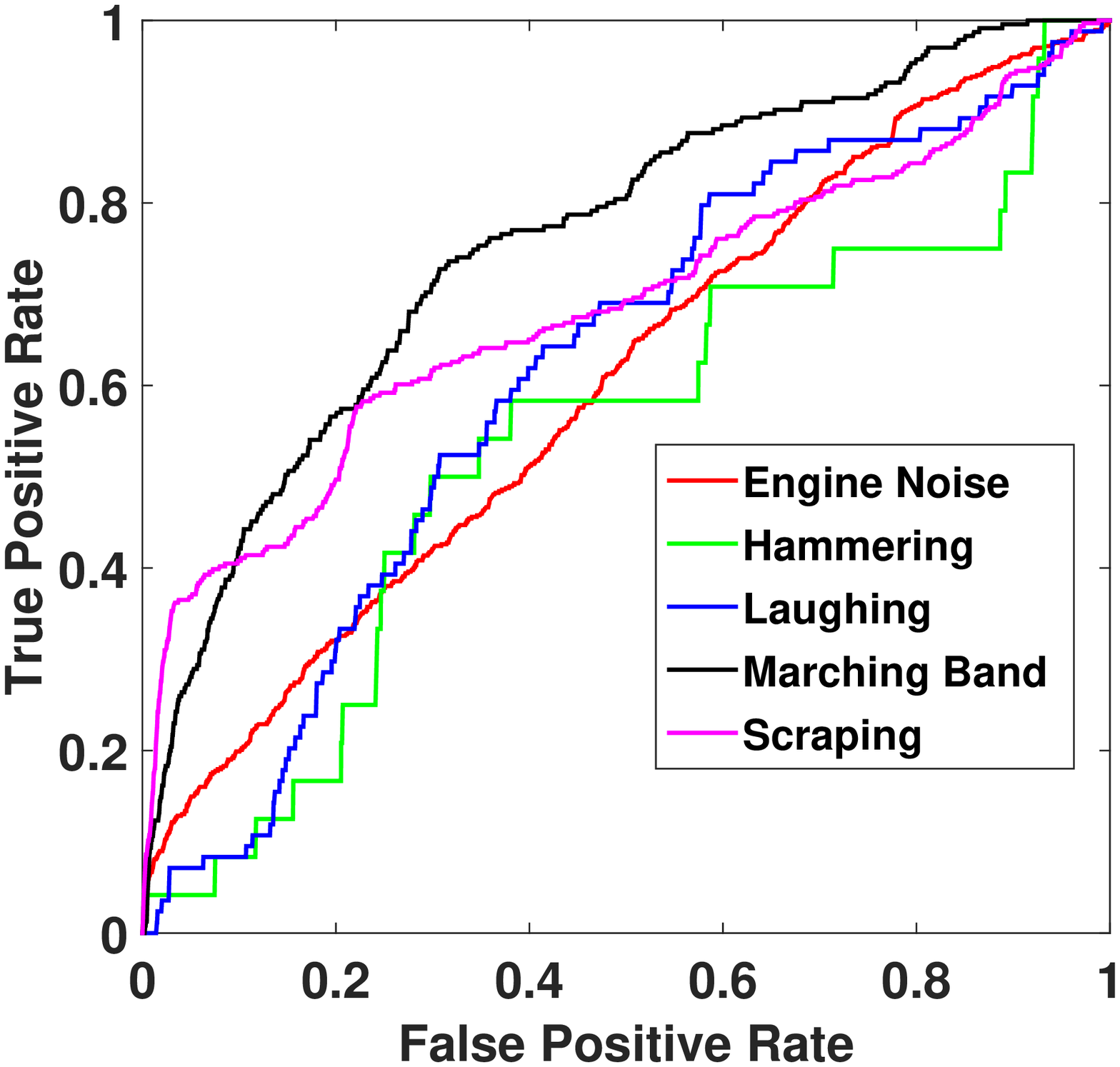}
\includegraphics[width=0.49\linewidth,height=1.5in]{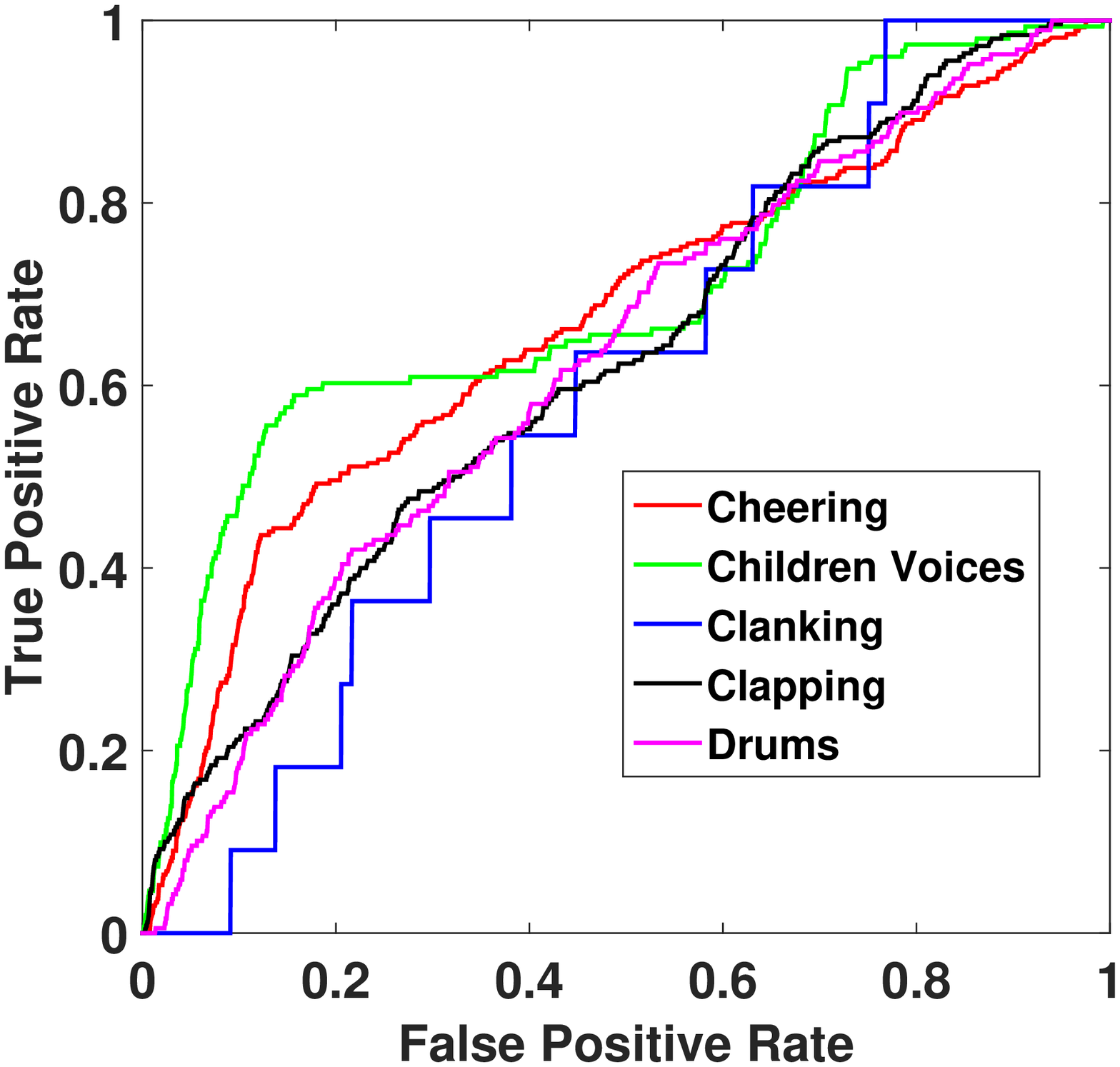}
\includegraphics[width=0.49\linewidth,height=1.5in]{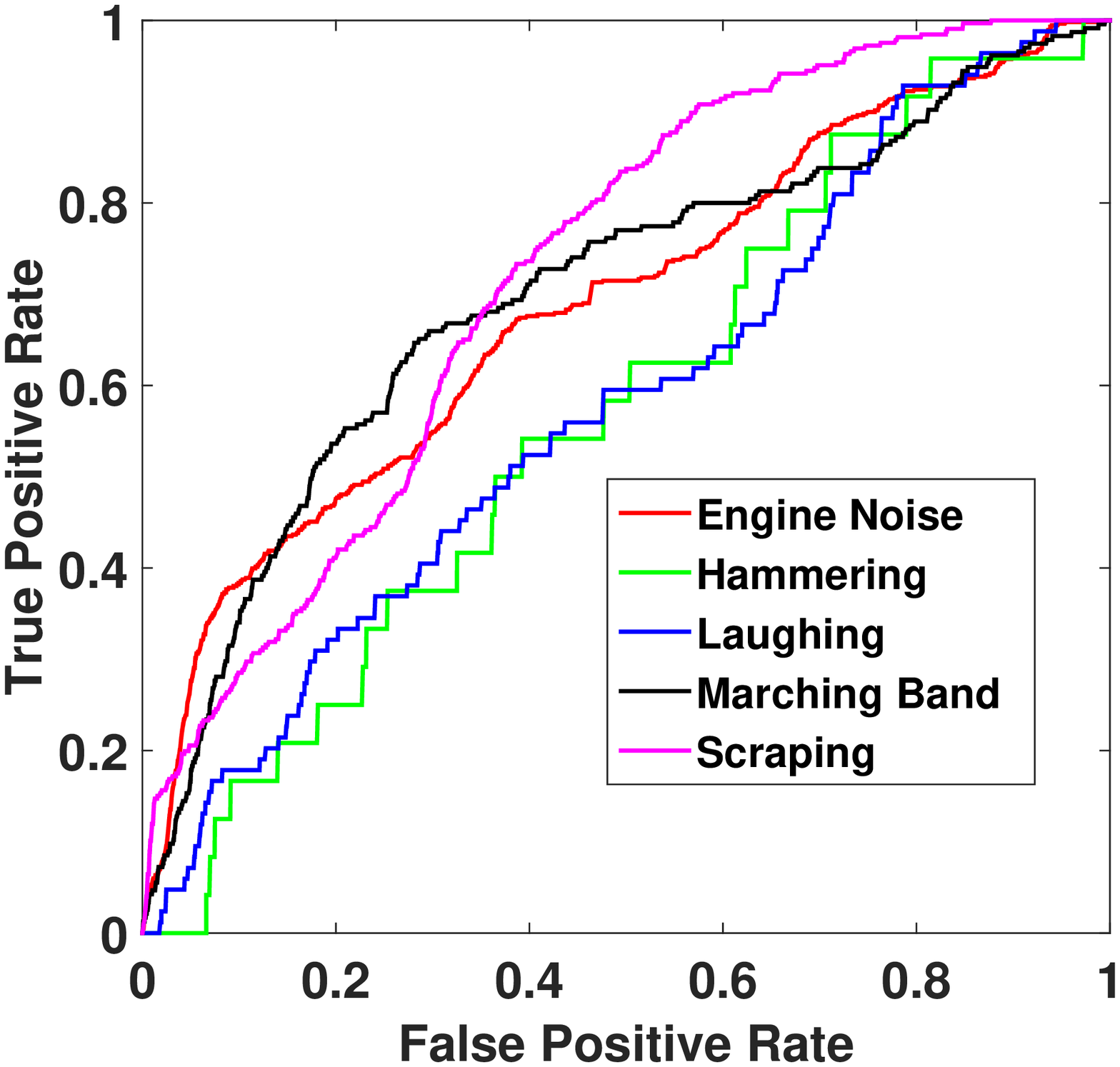}
\caption{ROC Curves for Temporal Localization (Top-miSVM, Bottom - BP-MIL)}
\label{fig:temploca}
\vspace{-0.15in}
\end{figure}

Once the ground truth labels with respect to an event have been obtained for all instances over all bags, we can analyze the performance in the usual fashion. We again present ROC curves and use AUC as the metric characterizing these curves. The best AUC values across all experiments for temporal localization using both \emph{miSVM} and \emph{BP-MIL} are shown in Table \ref{tab:tempauc}. The corresponding ROC curves are shown in Figure \ref{fig:temploca}. The figures in the upper row are for \emph{miSVM} and the figures in the bottom row are for BP-MIL. Compared to bag-level results, about $5\%$ drop in mean AUC is observed for both cases. For some events such as {\em Hammering} and {\em Laughing} the performance is poor for both frameworks. For others reasonable performance is obtained. Although these numbers are not exceptionally high, they are still significant since no temporal information was used during the training stage. Overall, AUC results  validate that our proposed framework can work for temporal localization as well. 
\begin{table}[t]
\centering
\caption{AUC for temporal localization of events}
\begin{tabular}{|c|c|c|}
\hline  
Events & AUC (miSVM) & AUC (BP-MIL)\\
\hline 
Cheering & 0.588 & 0.669 \\
\hline
Children's Voices & 0.665 & 0.705 \\
\hline
Clanking & 0.925 & 0.645\\
\hline 
Clapping & 0.585 & 0.626\\
\hline
Drums & 0.680 & 0.628\\
\hline
Engine Noise & 0.603 & 0.652 \\
\hline
Hammering & 0.542 & 0.572\\
\hline
Laughing & 0.548 & 0.581 \\
\hline
Marching Band & 0.758 & 0.701\\
\hline
Scraping & 0.684  & 0.724\\
\hline
\textbf{Mean} & \textbf{0.658} & \textbf{0.650}\\
\hline
\end{tabular}
\label{tab:tempauc}
\vspace{-0.15in}
\end{table}

\vspace{-0.15in}
\section{Discussions and Conclusion}
A framework for learning acoustic event detectors from  weakly labeled data has been presented in this paper. The learned detectors can both detect and temporally localize events in a recording. We show that we achieve reasonable performance for both tasks. Specifically, events such as {\em Clanking}, {\em Scraping}, and {\em Children's Voices} are easy to detect using both SVM and neural network approaches. On the other hand events such as Drums, Hammering and Laughing are hard to detect using both methods. Large-scale implementation of both methods on a larger number of events might be able to give a better insights for both methods. The overall performance of both methods is similar. Given the limited amount of training data, a mean AUC of around $0.7$  demonstrates the success of the proposed approach. Larger datasets are expected to result in better performance for both cases.  

An important factor in this framework is the representation used to characterize the audio segments, {\em i.e.} the instances in the bags. We employed  Gaussian mixture based features ($\vec{F}$ and $\vec{M}$). We do not necessarily claim that these are the best features for audio event classification; nonetheless they have been shown effective for AED in short audio segments \cite{23}. Other features may be expected to result in improved overall MIL performance. 

The success of our method is an important step towards reducing the dependence on strongly labeled data for learning audio event detectors. It shows the pathway to utilize the vast amount of multimedia data available on the web for audio event detection. The weak labels for web multimedia (audio) data can be inferred automatically from associated metadata which can then be directly used in the proposed framework. An interesting and extremely useful product of our proposed framework is the ability to temporally locate events in the recording. This is significant since this information was not present in the original data in the first place. Moreover, the predicted instance level labels can be further used in an active learning framework to improve performance.  

A number of factors still need to be investigated to create a state-of-art acoustic event detection mechanism in which learning is done using weakly labeled data. We need to consider a large set of events for a more comprehensive view of our framework. Moreover, investigations are required into more effective multiple instance learning methods which can help improve the overall performance and scalability of framework. Multimedia event detection requires detection of higher concepts. In the current work we focused on detecting finer events and their detection can be used for detecting higher concepts. Higher-level concepts in turn may be used to guide MIL. All of these are directions of currently ongoing and future work.

\bibliographystyle{abbrv}
\bibliography{references}  
%
%

\end{document}